# Second Product Line Practice Workshop Report

Len Bass
Gary Chastek
Paul Clements
Linda Northrop
Dennis Smith
James Withey

*April 1998*

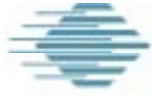

**Carnegie Mellon**
**Software Engineering Institute**

Pittsburgh, PA 15213-3890

# Second Product Line Practice Workshop Report

Len Bass
Gary Chastek
Paul Clements
Linda Northrop
Dennis Smith
James Withey

*April 1998*

**Product Line Systems Program**

# Table of Contents

# List of Figures

# Abstract

The second Software Engineering Institute Product Line Practice Workshop was a hands-on meeting held in November 1997 to share industry practices in software product lines and to explore the technical and non-technical issues involved. This report synthesizes the workshop presentations and discussions, which identified factors involved in product line practices and analyzed issues in the areas of software engineering, technical management, and enterprise management.

# 1  Introduction

## 1.1  Why Product Line Practice?

Historically, software engineers have designed software systems for functionality and performance. A single system mentality prevailed. Little attention was paid to the consequences of a design in the production of multiple software-intensive products. Large software development, acquisition, and reengineering efforts undertaken with this single system mentality perpetuate a pattern of large investment, long product cycles, system integration problems, and lack of predictable quality. Each product involves vast investments in requirements analysis, architecture and design, documentation, prototyping, process and method definition, tools, training, implementation, and testing with little carried forward to future products.

Many organizations have realized that they can no longer afford to develop or acquire multiple software products one product at a time. To retain market share in the global economy, they are pressured to introduce new products and add functionality to existing ones at a rapid pace. They have instead adopted a product line approach that uses software assets to modify, assemble, instantiate, or generate multiple products referred to as a product line.

A product line is defined as a group of products sharing a common, managed set of features that satisfy specific needs of a selected market or mission. It is most economical to build a software product line as a product family, where a product family is a group of systems built from a common set of assets.[1] In fact, the products in a software product line can best be leveraged when they share a common architecture that is used to structure components from which the products are built. This common software architecture[2] capitalizes on commonalities in the implementation of the line of products, and provides the structural robustness that makes the derivation of software products from software assets economically viable. The architecture and components are central to the set of core assets used to construct and evolve the products in the product line. When we refer to a product line, we always mean a software product line built as a product family.

---

[1] A software asset is a description of a partial solution (such as a component or design document) or knowledge (such as a requirements database or test procedures) that engineers use to build or modify software products [Withey 96].

[2] A software architecture of a computing system is the structure or structures of the system that consist of software components, the externally visible properties of those components, and the relationships among them [Bass 98].

Product line practice is the systematic use of software assets to modify, assemble, instantiate, or generate the multiple products that constitute a product line. Product line practice involves strategic, large-grained reuse as a business enabler.

Some organizations refer to the core asset base that is reused on systems in a product line as a platform. Terminology is not nearly as important to us as the underlying concepts involved, namely, the use of a common asset base in the production of a set of related products.

Some organizations have already experienced considerable savings in using a product line approach for software system production. Other organizations are attracted to the idea but are in varying stages of integrating product line practices into their operations.

In January 1997, the Software Engineering Institute (SEI) launched a technical initiative, the Product Line Practice Initiative, to help facilitate and accelerate the transition to sound software engineering practices using a product line approach. The goal of this initiative is to provide organizations with an integrated business and technical approach to systematic reuse of software assets so that they can more efficiently produce and maintain similar systems of predictable quality at lower cost.

One of the strategies to reach this goal involves direct interaction with and nurturing of the community interested in product line practice. This transition strategy has been executed in part by a series of product line workshops organized by the SEI. The workshop described in this report is the second SEI workshop to bring together international groups of leading practitioners from industry to codify industry-wide best practices in product lines. The results of the first such workshop are documented in an SEI report entitled *Product Line Practice Workshop Report* [Bass 97]. The SEI has also refined the workshop results through work with collaboration partners, participation in other workshops, and continued research. In addition, the SEI is producing a framework[3] for product line practice. The framework identifies the essential elements and practices that an organization should master for successful deployment of a product line. The framework categorizes product line practices according to software engineering, technical management, and enterprise management. These categories do not represent job titles, but rather disciplines. The framework is a living document that will grow and evolve.

## 1.2 About the Workshop

To connect with the product line practice community, learn the practices and issues in current industry and approaches to software product lines, and obtain feedback from experts on the first draft of the SEI Product Line Practice Framework, the SEI held a two-day Product Line Practice Workshop in November 1997. The participants in this workshop were invited based upon our knowledge of each company's experience with strategic software reuse through software product lines. Together, we elucidated and discussed the issues that form the backbone of this report.

---

[3] An initial version of the SEI Product Line Practice Framework will be accessible to the public from the SEI Web site in late 1998.

The workshop participants included

- Len Bass, Product Line Systems Program, SEI

- Staffan Blau, Ericsson

- John Brady, ALLTEL

- Gary Chastek, Product Line Systems Program, SEI

- Paul Clements, Product Line Systems Program, SEI

- Sholom Cohen, Product Line Systems Program, SEI

- John Curtis, Lucent Technologies

- Pat Donohoe, Product Line Systems Program, SEI

- Helmut Korp, Motorola

- Emil Jandourek, Hewlett-Packard

- Linda Northrop, Manager, Product Line Systems Program, SEI

- Dennis Smith, Product Line Systems Program, SEI

- Scott Tilley, Product Line Systems Program, SEI

- Theo von Bombard, Bosch

- James Withey, Product Line Systems Program, SEI

- Peter Wood, Nokia

The participants had product line experience in the following domains: telecommunications, financial information systems, automotive parts, aircraft control, and consumer electronics.

Each guest at the workshop was asked on the first day to make a presentation explaining his organization's approach to developing software product lines. Before the workshop, the draft of the SEI's Product Line Practice Framework was sent to participants to provide a common structure for the presentations. This framework describes areas of expertise and practice that are essential for successfully developing, deploying, and maintaining a software product line. Participants were asked to comment on the framework as part of their presentations.

On the second day, participant presentations were summarized, then the participants divided into three working groups to explore the practices in software engineering, technical management, and enterprise management further. The working groups then presented their results to the entire group.

## 1.3 About This Report

This report summarizes the presentations and discussions at the workshop. As such, the report is written primarily for product line champions who are already working or initiating product lines practices in their own organizations. Technical software managers should also benefit from the information.

The report is organized into four main sections that parallel the workshop format:

1. Introduction

2. Product Line Experiences: Summary of Participants' Presentations

3. Product Line Practices and Issues: Working Group Reports

4. Summary

The section following this introduction, Product Line Experiences: Summary of Participants' Presentations, synthesizes the product line experience of the workshop participants by describing the contextual factors and the software engineering, technical management, and enterprise management practices and issues. Section 3 is composed of the three working group reports on selected practices and issues in software engineering, technical management, and enterprise management, respectively. The summary in Section 4 recaps the major themes and suggests future directions. Additionally, a glossary of terms is provided.

# 2 Product Line Experiences: Summary of Participants' Presentations

Before the workshop, each participant received a draft of the SEI's Product Line Practice Framework to provide a common presentation structure. The Framework describes areas of expertise and practice that are essential for successfully developing, deploying, and maintaining a software product line, and participants were asked to comment on the framework as part of their presentation.

The purpose of this workshop was not to catalog individual product line approaches, but rather to synthesize the combined experiences of many organizations to help the SEI build and improve our model of product line practice, as expressed in the framework. Following the tradition of the first workshop report [Bass 97], this section summarizes the presentations as a group, rather than individually, by re-casting the presentation in terms of a set of common themes that applies across all presentations. These themes are contextual factors, and (following the partitioning of issues in our framework) software engineering, technical management, and enterprise management issues. We address each theme in turn.

## 2.1 Contextual Factors

Contextual factors describe the environment in which the organization exists or existed when it launched the product line effort; this includes a description of its goals, the technical assets in place, its business state, and how the organization is/was situated in its marketplace.

**Motivation.** The participants in this workshop reflected broad experience in successful product line strategy. One of the common themes they expressed was employing the product line strategy as an approach to achieve large-scale productivity gains (rising by a factor of four in one case) and time-to-market improvements (reduced by a factor of four in another case). As in the first workshop, there was an underlying sentiment that product lines were not just a good idea, they were essential to the organization's continued health in a market. To quote one participant, "We couldn't compete in the domain otherwise." Interestingly, several of the organizations moved to product lines not as a response to dwindling business, but to unprecedented growth. Maintaining market presence and sustaining that growth required "fully featured software" as the key to retaining competitive advantage. Without a product line strategy, hiring requirements would have been prohibitive. One organization projected a need to produce six times as much software (measured in terms of available features) over a four-year growth period. Increasing the staff size by six times was out of the question; at best, only a 10% growth per year in staff could be projected, growing the organization by 50% over the four-year period. Producing six times the products with only 1.5 times the staff precipitated a factor-of-four productivity growth goal for that organization, which they achieved in only three years using the product line approach. Other organizations expressed their goals in terms of reuse levels gained by using what they referred to as the platform, or core software base that is reused across systems in the product line. One organization reported an 80% reuse goal, while a second organization reported that their platform accounted for 89% (on average) of each product they marketed in the product line.

**Product line maturity.** Our participants represented a cross-section of industry experience in terms of how long they had been developing systems using product line approaches. One participant reported a 15-year background in building "product-line-like" systems, while others were in the early stages of implementing such an approach, with quantitative improvement results not yet in. Interestingly, none of the "latecomers" expressed any hesitation or misgivings about the approach they were just beginning to use, and all reported at least qualitative improvements. A couple of the participants were just beginning a product line in a new domain, but had previous product line experience in a different application area. Support for the approach was, at this workshop, not a function of how long one had been at it.

**Availability of legacy assets.** Our participants varied with respect to whether they developed their product lines from existing software assets or whether they had so-called "green field" (start from scratch) efforts. However, one legacy asset that all organizations had in common was long and deep experience in the domain of the systems. This validates the experience of the first workshop, when detailed domain experience was identified as an indispensable prerequisite for product line development.

## 2.2 Software Engineering

Although one of the lessons we have learned about product line development is that business and organizational factors are at least as critical to understand and manage as technical issues, building a product line is at its heart a software engineering task. Critical software engineering technologies that come into play during product line development include requirements management, domain analysis, architecture development and evaluation, exploitation of existing assets, component development, and testing. Of these, the following received particular attention during the presentations.

**Domain analysis.** Only one participant reported performing an explicit domain analysis step, even though product line engineering is based upon the well-known domain-engineering/application-engineering dual-life-cycle model. Most organizations tend not to have explicit, written domain models; rather, the models are often intuitively known by the people engineering the products (and the architectures for those products) that constitute the product line. Typical of this camp was one organization that reported that detailed requirements for the product line came from the single driving product that launched it, with the confidence and expectation that it would meet the needs of other projects as well. The organization that did perform the explicit domain analysis reported the following:

- The domain analysis effort took about three months.

- The domain analysis and development of a domain architecture together took six months, "which was a long time to keep management interested."

- They used a requirements-based domain analysis derived from the work done by Guillermo Arango.

- The process steps included setting the scope of the domain, gathering information, identifying and classifying features, performing a commonality analysis, performing a competitive analysis, and validating/evaluating the results. Features were represented using an object model.

- Artifacts produced included product histories, requirements inventories, a domain dictionary, a feature model, and a commonality.

- The domain analysis revealed that two previously separated parts of their application domain were in fact quite similar, with "no justification for different architectures or different development groups," contrary to previous assumptions in that community.

The participant concluded by reaffirming that "domain expertise is the most important thing." For most organizations that omit the explicit domain analysis step, however, there is a plaguing question: How does the organization protect itself against personnel turnover if the domain expertise resides only in the heads of the staff?

**Architecture development and evaluation.** As in the first workshop, architecture and architectural concepts played a critical role in the successful development of product lines. Nearly every presenter showed the architecture for his organization's product lines in the first few slides of his presentation. One organization separates the task of designing the architectures for systems into its own business unit. Interestingly, however, the creation of that architecture was not mentioned as a difficult problem. It is either the case that the organizations represented in this workshop all employ high-caliber architects who can design product line systems so productively and successfully that the process looks easy; the state of architectural practice is maturing to the point where standard architectures or architectural patterns, off-the-shelf components, and domain-wide reference models are making the process straightforward; or an in-depth discussion of architecture was viewed as too proprietary. As at the first workshop, the layered architecture was the most oft-employed architectural style, because of the portability across platforms it provides by separating platform-dependent/application-independent software components from platform-independent/application-dependent components. The layered view of an architecture is also the most useful to present to a large range of audiences. One organization's architecture featured components targeted to large-grained reuse, with the following properties:

- Each component provides a subset of the system's features.

- Each component is designed to be customized to specific applications, with pre-defined evolution points or "hooks."

- A component is "configured, packageable, distributable in a stand-alone fashion."

In short, these components represent large, functionally significant assets, which is in keeping with a consistent theme we often observe with successful product lines. This approach leads to a paradigm for system-building that emphasizes composition instead of generation (let alone programming).

None of the participants mentioned a separate architecture evaluation step, though two of the organizations do have software architecture evaluation practices.

**Exploitation of existing assets.** For those organizations that did make extensive use of pre-existing assets to build their product line, no particular methodologies or technical approaches were employed. Rather, ad hoc inventory and reengineering methods were used. One organization reported that they "have found it easier to take an evolutionary approach and roll an existing asset into a platform [making it more generic]." This organization had a from-scratch effort fail (due to poor stakeholder communication, it was thought), so building from a legacy base enjoyed greater support from management. A second organization tries to re-build (rather than build anew) whenever possible because they feel that "green-field" efforts lead to low initial quality and have a much more problematic critical path. Yet another organization characterized their product line as the "migration of existing solutions to a larger group."

## 2.3 Technical Management

Technical management includes those management practices that are directly related to maintaining a healthy project. They include metrics and data collection/tracking, configuration management, and planning.

**Metrics and data collection/tracking.** Developing a product line can entail a significant up-front cost, as shown by Bass and Brownsword [Bass 97, Brownsword 96]. In many cases, the investment required can be a daunting influence. Management must approve re-organizations, re-engineer the enterprise's interface with its customer base, install aggressive training programs, and in general inflict a massive cultural shift onto the organization. The product line champion or advocate must be able to show that the changes are paying off. In this workshop, several of our participants had quite sophisticated ways to measure the improvement, beyond the time-to-market and lines-of-code productivity measures mentioned at the first workshop and again at this one. (Once again, as at the first workshop, cost was not considered an important measure of improvement when compared to measures of market re-sponsiveness.) One participant reported that to measure product line productivity over a three-year period, his organization tracked

- the number of products shipped (increased by a factor of five)
- the feature density of products shipped (increased by a factor of four)
- product volume shipped (up by 18 times)
- number of new features released per year (up by three times)
- product volume shipped per person (up by four times)

**Configuration management (CM).** CM is a crucial capability for maintaining a product line. Products in the product line differ from each other, of course, but different versions of the same product also differ from each other. Participants strongly agreed that being able to quickly and reliably rebuild a particular version of a particular product (when the components may have undergone modifications since that product/version was originally released) was the defining problem of product line CM. Part of being able to achieve this capability is structuring the products such that long recompiles are not necessary; more than one participant mentioned that the field support requirements of their organization would not allow rebuilds that took even a few hours. Interestingly, nearly all of the participants mentioned that they used the same commercial configuration management product. The consensus about this product was that it is adequate for the job (whereas simple version-control tools such as the software change and configuration control system [SCCS] are not), but it is not usable right out of the box. Rather, careful attention must be paid to setting up and maintaining the appropriate information structures that will allow the rebuild requirements to be met.

One organization proceeded as follows: One directory is assigned per architectural component (or per designer, which is often the same thing). Every component is assigned a label by which one can tell what components/versions have been used to test it. Then, scripts are created that produce particular versions of a system, an architectural layer, or some other pre-integrated "chunk" of the system. This particular organization tests each architectural layer as a whole, and then builds a system out of three to four tested layers. (Again, we see the theme of a system built from a small number of large pieces.) The emphatic advice of this participant was that the directory structures, the policies, the build levels, the kinds of builds, and so forth must be decided by the products group and not "some detached Tools and Methods Department" whose stake in the game is once removed at best.

**Planning.** Planning was not explicitly addressed in the presentations, although there was firm consensus that planning is critical to product line development.

## 2.4 Enterprise Management

Enterprise management is the term we use for the management of the business issues that are visible at the enterprise level, as opposed to just the project level. Enterprise management includes those practices necessary to position the enterprise to take fullest advantage of the product line capability. It includes achieving the right organizational structure to build a product line, ensuring proactive management, building and maintaining appropriate skill levels, managing the organization's customer interface, facilitating efficient intergroup collaboration, and performing the necessary business analysis to make sound financial and planning decisions. Of these issues, two received significant attention in the presenters' talks: organizational structure and business analysis.

**Organizational structure.** Previously, our model of a product line organization had been that one group builds the core asset base or "platform," while a separate group (or groups) builds the products in the product line for delivery to customers. This was the model that emerged from the first workshop. The rationale for this model is that producers of individual products will tend to have the interests of their particular products at heart, whereas members of a generic asset unit will be less parochial and produce more honestly generic assets. One participant at this workshop, however, argued for the reverse. The problem with separating the product and core asset groups, he maintained, is that the goal is not to have a beautifully engineered core asset base, but a core asset base that helps the enterprise make money. A separate asset development group may be more likely to produce beauty, not profit, and the separation inhibits the intimate feedback that is necessary to make the assets usable across the line of products. In this person's view, someone responsible for the profit and loss of the business unit must decide what to make generic and when to let products go their separate ways. Separating the groups makes more sense, he argued, in start-from-scratch product line efforts, assuming that the issue of how to fund the core asset development can be resolved. A compromise is to rotate people between the two groups. The key is to have someone with specific oversight responsibility and authority mandate the construction of generic assets. It is not necessary, though, that the generic asset be built by a separate generic asset group.

It may be the case (this hypothesis was not tested at the workshop) that to separate or not to separate the asset development group depends upon how much development must be done to produce a product from the platform. If little work is involved, then it makes sense to have most people work in a dedicated fashion to produce the core assets. If turning out a product entails significant development effort, then it makes more sense to have dedicated product groups with less emphasis on the core assets.

A second organization reported that they maintained a steering committee whose charter was to decide when to fold new, common assets into the product line asset base to make it more generic for future development.

**Business analysis.** This area covers a range of topics, but the presenters focused on two: funding the production of the generic product line assets and using product lines as a springboard for the enterprise to enter a new business area.

Funding the development of generic product line assets was, according to several participants, a critical issue. In a from-scratch effort, the generic assets have to be created and paid for. Somehow the cost has to be amortized across more than the first project or two that will use the assets. In a reengineering effort, the effort to make specific components generic also must be paid for. If core asset or platform group is separate, a policy for funding it must be selected. Product revenue is one source, and some of our participants used this approach; the mechanism was a "tax" on the product groups to pay for the use and development of core assets. Another source is from the enterprise's research and development (R&D) budget, which was the approach taken by another of our participants.

In terms of entering a new business area, more than one presenter made the following point, and it resonates well with other product line experience reports we have seen. Alphanumeric pagers beget digital pagers. A command-and-control system product line provides an entry into the air traffic control market. An air traffic control product line facilitates entry into the marine vessel control domain, etc. Beyond the potential for electrifying productivity gains with a particular domain, allowing entry into entirely new markets may be the most dramatic enterprise effect brought about by a product line capability. This impact is usually not anticipated when the original product line is developed.

## 2.5 Summary

The presentations covered a broad range of topics; these topics all fit relatively cleanly into the partitioning proposed by our framework, which gives us confidence that the framework partitioning is a reasonable one.

We close this section by reporting on the list of "hard" issues that two of the presenters included in their presentations. Some of these will be addressed in the working groups' efforts to be reported on in the next section; others, however, serve as focal points for future community work. We have partitioned the issues into our framework categories, but left the wording as expressed by the participants. Notice that the bulk of the issues reside in the enterprise management category.

### Software Engineering Issues

- achieving reliability in the face of a test-case explosion for complex systems

- reengineering: when to re-architect the system

### Technical Management Issues

- metrics: which ones to collect and why

- lineage: traceability of problems, adaptability, and associated cm issues

- managing changes to the core assets: push or pull?

### Enterprise Management Issues

- managing customer lifetime support: making customers understand that the components in their long-lived systems will be obsolete in five years and the cost of spares will be prohibitive if they do not upgrade

- warranty: how to bid and define

- customer confidence in maturity of the product line, especially in a safety-critical application; customer training

- funding models: a "major, major" issue
- constancy of management purpose
- constancy of organizational direction
- local optimization issues: time to market, technology
- long-term ownership and support
- platform must span product lines to get its funding from several sources
- platform must quickly (six months or less) generate visible results

# 3 Product Line Practices and Issues: Working Group Reports

## 3.1 Software Engineering

This working group discussed two of the software engineering practice areas important in a product line approach: mining assets and domain analysis.

### 3.1.1 Mining Assets

Four steps were identified in the process of mining existing assets to build or to augment the set of core assets:

1. Decide on commonalities among existing components or on the need for generic components.

2. Decide that mining is the correct mechanism for achieving a new core asset.

3. Create the generic component.

4. Install the generic component in the asset base for adoption by users of core assets.

We now discuss these steps in more detail.

**Decide on commonalities**. The decision that commonalities among products exist but are not being addressed in the asset base may come from several sources. One source is a collection of product managers who observe commonalities in their collective products. This can result from a strategic decision to gain competitive advantage from a new and more efficient capability of being able to configure new product offerings more rapidly. The realization often occurs when a new product is about to be constructed. A second source of the need for commonality to be captured in the asset base is the sales force who observe features in competitors' products or who hear the need for features in their own products from customers. A final source that pushes the decision to generate a new generic component is the passage of time and the arrival of new products in the marketplace. In this latter case, the new component may be a layer that hides details of the new version or that provides a virtual layer across several different products offering basically the same service.

**Decide to mine.** Once the commonalities have been identified at a high level, a decision is made as to whether these commonalities should be added to the asset base through the mining of an existing asset or through the creation of a new generic component from scratch. The use of existing assets over "green-field" efforts is the norm.

This step and the prior step are often interrelated. That is, the decision to embrace features added to support a particular customer in a particular application may trigger a decision to add a generic component for that feature to the asset base.

**Create generic component.** Determining the features that the generic component will have is often the most contentious portion of the process. Different organizations approach this in different ways. Some organizations gather all of the stakeholders together (generally representatives of the affected products and the sales force) and have them decide on the features of the new component based on customer requirements. Others make decisions based on potentials from the existing asset base. Another approach is to decide on a potential market niche that is currently unfilled.

Once the features have been identified, then the new component can be created. A portion of the process of determining the features is to decide on what base the new component will be created. One organization has the group that created the original component create the generic one. This organization then ensures that the generic component will operate in the environment from which it was mined and modified. After this validation, it installs the component in the asset base.

**Install new component in asset base.** Once the new generic component has been installed in the asset base, other groups treat it as any other component in the asset base. That is, new releases of the asset base are made and the product groups decide what components they will use from the new release. It is crucial for the assets to be kept under strong configuration management controls.

## 3.1.2 Domain Analysis

There are a variety of situations when domain analysis can be effective, as described below.

Domain analysis can be performed before the construction of the product line as the initial step in asset development. The domain analysis defines the scope of the target products and identifies key common features and their variations across current, future, and competitor systems. The domain analysis results feed the architecture and framework development process.

Domain analysis can also be performed piecemeal during the development of product lines. With this approach, the domain analysis is performed in a bottom-up fashion covering small portions of the product line and expanding in scope over time.

The domain analysis can also begin with an existing system that is representative of the domain. When a new product is desired, such as when a new contract is acquired, a domain analysis is performed to determine what is usable from existing systems and how to incorporate new features.

Industry practice suggests that the domain analysis step can be omitted when there is deep expertise in the problem domain and systems for the domain.

A number of observations were made about domain analysis:

- The process is time consuming. It involves senior people, and it takes a long time to get all of the details correct.

- It is important that the domain analysis is perceived as having a direct link to delivered systems. Otherwise, managers and customers can get impatient and view the effort as not having value.

- Domain analysis is a useful training tool. It helps the marketing staff to understand the differentiation among the products, and it helps the engineering staff during development.

- The interaction between domain analysis and architecture development is not well understood. Some follow a waterfall model, where the domain analysis step is completed before architecture work begins, while others believe a cyclic approach should be followed. The waterfall model is unrealistic in most situations.

- Horizontal domain analysis (understanding the relationship among different features that provide different services) has different considerations from vertical domain analysis (understanding the relationship among different layers that combine to form a usable collection of products).

## 3.2 Technical Management

The technical management working group discussed practices and issues related to the management of an organization's technical product line practices. While these practices occur at the software engineering level, the orchestration of these practices occurs at the technical management level.

The discussions focused on the following product line practice areas: metrics, testing, configuration management, and planning. In this section we summarize the working group discussions by practice area, then present the identified areas for further work.

### 3.2.1 Metrics and Data Collection

The purpose of metrics and data collection are to validate that the products, and the processes and procedures used to create and maintain those products, meet the business goals of the organization. Metrics measure what a manager needs to know to lead and track the organization's technical effort.

The metrics discussion was focused by the business goals that the metrics address: productivity, product time to market, and product quality.

#### 3.2.1.1 Productivity

Productivity can be defined for either an individual engineer or the organization as a whole. At the engineer level, productivity is typically defined as the number of lines of code produced by an engineer per year. At the organization level, productivity is the number of different types of products shipped by that organization per year.

The group felt that both of these measures of productivity should be weighted by the complexity of the products produced, as captured by the number of features in those products. Product features are important because they are what the customer ultimately sees. For a product line, features are counted across all the products in the product line.

These productivity metrics are not different for product lines; however, we would expect these metrics to capture the improved productivity that presumably results from the product line approach.

### 3.2.1.2 Product Time to Market

Product time to market was defined as the time from the initial conception of a product to the delivery of that product. It includes the time to define and determine the feasibility of the product as well as the time to specify, design, implement, test, and deliver the product.

Determining the precise starting time for a product, however, is difficult. One participant suggested a "retrospective" definition of when a project starts: once five percent of the projected engineering resources have been spent, the project has begun. This definition fits well with an incremental or spiral development view.

While the ability to produce a product quickly (i.e., a short time to market) is a strong market advantage, the use of that ability must be tempered. It is more important to deliver a product at the right time than to deliver a product quickly. Too frequent releases may saturate the market: the time to product release must be less than the market absorption time.

Other product line-specific metrics mentioned include the platform time to market, the average product time to market, and the feature time to market.

### 3.2.1.3 Product Quality

Product quality was defined as the number of defects in a product. The group discussed the expected effect of a product line approach on product quality and other quality-related metrics.

The number of expected defects per product should be smaller with a product line than with a single system due to reuse of the previously tested code and platform. The time to resolve a customer problem and the effort to fix a defect probably stays the same for a product line. However, the cost of the fix can be amortized over all the products in the product line, reducing the per-product cost of that fix. This could be the basis for a "fix effectiveness" metric.

The group hypothesized that the probability of a defect would decrease as platform use increased, and that the initial (i.e., at first delivery) probability would be high regardless of whether the product was a single system or part of a product line.

### 3.2.1.4 Conformance to the Product Line Architecture

While not strictly a metric, conformance to the product line architecture is an important concept for a product line.

The conformance of the products within a product line to the product line architecture is a measure of the commonality, compatibility, consistency, and cohesiveness of those products. Since the reference architecture captures the goals of an organization, conformance to that architecture is also a measure of how well that product line meets the company goals-in short, a measure of the success of the product line.

## 3.2.2 Testing

Test effectiveness is closely related to product quality: more effective testing should yield higher quality products. The group's discussions centered on the relationships between product quality and product testing in a product line.

The group made two key testing-related observations about product lines:

- The core assets of a product line are tested early and often.

- The testing effort for a product line (i.e., the number of times the test cases are applied to the code) is greater than the effort for a single system. This results from the need to prevent erroneous fixes from adversely affecting multiple products and the multiple targets that are typically found in a product line.

The working group drew the following conclusions from these observations:

- Core asset quality should consistently improve from the first to the later releases.

- Although the per-component testing increases, the per-product testing for a product line decreases due to the reuse of previously tested code. Also, the cost of the increased testing effort can be amortized over all of the products in the product line.

- There should be more product bugs than core asset bugs as the product line matures.

The final observation led to the conjecture that the ratio of the number of core asset bugs to the number of product bugs found during system testing may be a metric that captures core asset quality. It was further suggested that the ratio should be scaled by the core assets' relative contribution to the product.

## 3.2.3 Configuration Management

Configuration management was viewed as a way to control the building of a product. Source code, system requirements, and test cases should be controlled by the configuration management system.

The required services of a configuration management system include

- version management and branching

- labeling and control of labeling

- storage and control of historical information to re-create previous versions of products

- easy and rapid re-creation of previous configurations

- mapping of component versions to product versions

- rapid access to any configuration, including the testing and support environments (e.g., compiler, operating system)

Product lines do not change how one does configuration management; however, the content controlled by the configuration management system is different for product lines, requiring the ability to

- record product context. The product context is a snapshot of the complete product-development environment for that product, including information about the compilers and debuggers used. It also depends on a mapping from the components to the system versions. Product context is important because defect detection and correction dependends on the configurations targeted for the change.

- switch contexts from one product to another quickly. A product line will see a greater frequency of context switches than a single system.

- partition core asset artifacts from product-specific artifacts

The working group stressed the value of a change management policy for controlling the product line's core assets. The policy is an abstract set of guidelines that defines

- how and when to make a change

- the scope of a change

- when to incorporate a change into the rest of the product line

- when to isolate a product or platform

- how to deal with white-box architectures

Other related observations include the following:

- Each project should maintain a configuration management checklist enumerating the project's policies and procedures for developing new versions and branching. The checklist is in effect a "rules of the road" for artifacts.

- An organization must be at CMM[4] Level 2, at least with respect to configuration management, by the time the first product in a product line is shipped.

## 3.2.4 Planning

Planning for a product line is similar to planning for a single system: the problems addressed are primarily people (rather than technical) problems. However, product line planning is more critical because the product line dependencies are more critical.

---

[4] Registered in the U.S. Patent and Trademark Office.

The working group offered the following observations based on their experience:

- There is value in crossing the application and domain engineering staff.

- Do not make the manager of the core asset base also responsible for the product; core asset and product development should have different goals and rewards.

- The reward system needs to be considered very carefully. Bonuses and other rewards may have unforeseen negative effects.

- Very few know how to manage an architecture effort and understand when it is on track.

### 3.2.5 Issues for Further Investigation

The working group identified the following open product line issues related to technical management:

- how to measure the expected decreased probability of a defect as a result of a product line approach

- how to guarantee that a particular fix is in the spirit of the product line

- how to determine
    - the "right" tests for a product line
    - if a given test suite covers the product line
    - the boundary between what is a core asset and what is not

- how to manage an architecture effort

- how to predict, or at least plan for, likely future products and reflect that in the product-line architecture

- how to mitigate the risk associated with the determination of likely future product features

- how to plan for the business, technology, and process paradigm shifts associated with product lines

- need for better understanding of the product line process to permit better planning. In particular, a process model is needed for the required product line activities, captured in a "cookbook" of what needs to be done to build and deliver a product. This cookbook would capture what is unique about product line development.

## 3.3 Enterprise Management

This working group sought to describe product line practice from the perspective of the enterprise. Two practice areas were discussed: determining organizational structure and determining a product line production strategy.

### 3.3.1 Determining Organizational Structure: Architecture and Other Assets at Multiple Organization Levels

*Organizations which design systems are constrained to produce systems which are copies of the communication structures of these organizations* [Conway 68].

One of the significant enterprise challenges is determining the right organizational structure to implement a product line approach successfully. The working group explored the levels in the organization at which software architecture and other assets can be developed, and they described some intrinsic properties about these assets when developed at those levels. The working group then explored the interface of architectures and other assets to product development projects. The practices and organization of a product line approach seem to vary according to the amount of product variation that is incorporated in a software architecture and the software components.

With respect to software architecture, Conway's observation is a good rule of thumb: software products do tend to reflect the organization's communication structure [Conway 68]. The scope and content of the product architecture tend to be driven by the organizational structure and funding source. However, working group participants reported that business strategy determined the organizational level at which architecture and other assets are developed; and that design constraints embedded in the organizational structure do not occur by happenstance, but by conscious decisions to increase competitive advantage.

Working group participants reported that software architecture and other assets are typically developed at one or more levels in the organization: at the corporate level, the business unit level, or the product line organization level as shown in Figure 1. A business unit is a profit and loss center with a portfolio of related products and markets. For example, an office products business unit may consist of printer, photocopier, and computer product lines. The working group considered a product line organization to be an organizational entity responsible for related products in a specific market. To avoid confusion, we will also assume that a product line has one architecture and a set of assets.[5]

---

[5] A product line organization may be responsible for products offered in different market segments. If the range of functionality is significantly different between segments, a software architecture and other assets may be developed at a sub-product line level. For example, a product line organization may have one architecture and assets for the low-end home market, and another set for the high-end corporation market. If this is the case in your organization, then the discussion in this section pertaining to business units may relate to your "product line," and the discussion pertaining to product lines may relate to your "sub-product line" or product family.

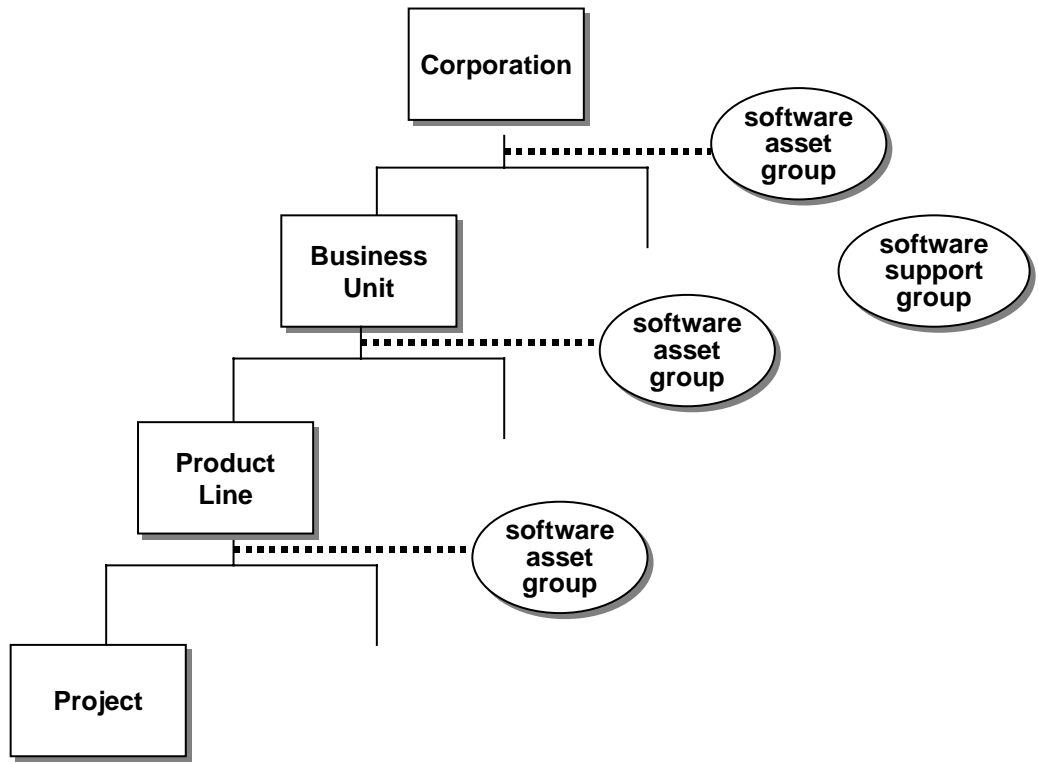

*Figure 1:    Asset Development Levels*

Regardless of level, two groups are usually involved: an architecture group and an asset or platform development group. (In Figure 1 both groups are represented by one oval labeled "software asset group.") The architecture group is usually led by a senior architect and composed of members from development projects. This group specifies the architecture and may develop training materials. The group may be permanent, producing refinements to the design, or ad-hoc, and either phased out or combined with the asset development group after a baseline architecture has been defined. The asset development group develops the assets: components, languages, tools, methods, procedures, and information systems that help product development groups. Frequently (as reported in the working group) they construct a "platform," a standard configuration of components or asset base used in each development project. The software asset group usually maintains this platform, testing in-house or commercial components and issuing periodic releases with updates. Also mentioned by working group participants was the existence of a technical support group that provides training and consulting services often on a contract basis. This group is shown in Figure 1.

The scope and content of the architecture and other assets appear to differ at each organizational level. The role of the asset group in the organization, specifically its interaction with product development projects, also varies by organizational level. We will now look at each level in more detail.

### 3.3.1.1 Corporate Level

Assets developed at the corporate level are viewed as part of a strategic competency that must be nurtured to retain and increase market share. The architecture and other assets are usually based on protected technology that underlies a large number of products in multiple businesses. Three examples are a domain-specific specification language and toolkit, a real-time point-of-sale inventory subsystem, and a check-processing subsystem. The assets are shared across profit and loss centers (business units). Initially funded with corporate research and development money, the group, once successful, either delivers the assets to asset groups in the business units or is turned into a separate for-profit organization providing products and services to the open market. In the latter case, special transfer pricing agreements for assets used by a product line may be arranged with the parent organization. The decision to spin-off the asset group depends on how the technology is protected and on the projected economies of scope. As a rule, if the variety of products using the software assets is larger in the marketplace than in the parent organization's business units, then a new organization should be formed. This has two side benefits: competitors pay a price premium to enter the market (competitors are using your assets) and lock-in of obsolescent technology is less likely (market forces will sustain innovation).

### 3.3.1.2 Business Unit Level

Architecture specifications and components developed at the business-unit level (across product lines) typically include system services such as communication primitives, operating system components, and hardware interface subsystems. Often the architecture and components form the computer system platforms used by the product lines.

The scope and type of software assets developed by this group are influenced by the funding model used by the business unit. If the asset group is funded by taxing the product line organizations, then ultimately only common software assets will be developed. Because product line managers view the tax as a loss of personnel, they support the asset group to the extent that it delivers products and services that meet their particular needs. To maintain management support, the manager of the asset group therefore develops software assets that have appeal to the greatest number of product lines in the business unit. Assets that have been developed or optimized for a specific product line tend to migrate under that group.

The tax-funding model also changes expectations of the products and services provided by the asset group. Product managers expect off-the-shelf components: bug-free, plug-and-play software. They do not see the need for the asset group to charge for component integration or customization. Co-development of components and other assets-an effective means to achieve optimum performance-does not frequently occur.

If the asset group is funded by research and development funds, then the manager has the flexibility to develop product line-specific components and tools and engage in collaborative funding arrangements. The asset group can develop software assets that further the strategic goals of the business unit, even if many product lines in the business unit do not directly benefit from this activity. Under this funding model, product line managers regard collabora-tion as a way to leverage their development capabilities. However, this funding model requires more management direction; without it, the asset group may become an ivory tower, not motivated to service the product line groups in the short term. In some organizations, a strategic steering committee composed of product line managers directs the development funds of the asset group.

### 3.3.1.3 Product Line Level

Assets at the product line level typically include large-grained subsystems and testing tools. Because they can contain more application-specific components, product line platforms usually have higher leverage than platforms developed at the business unit level. Less work is needed to develop a specific product. However, they are usable across a smaller number of products. The tradeoff is shown in Figure 2. The y-axis, "product features," indicates the number of different product features that are included in the platform. Under controlled circumstances, average percent-reuse figures may be substituted for the product feature metric.

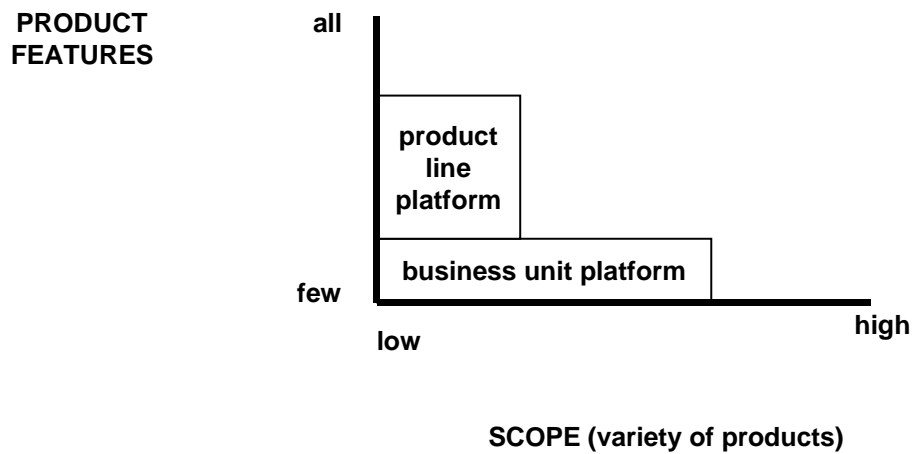

Figure 2:   Tradeoff of Scope and Product Features in Different Platforms

## 3.3.2 Determining Product Line Production Strategy

Product line managers make the decision to develop software assets for a product line based on an analysis of the business: the competition, future demand for new product features, current capacity, technology maturity, and possible pricing strategies. An outcome of the analysis is a plan for future products. This plan drives resource allocation decisions that the product line manager has to make, such as: "Do I allocate engineers to build software assets or not?"

Implied in resource allocation decisions is the selection of a specific production strategy. A production strategy is the engineering approach for building the products in a product line. It describes the actors, the different products they produce, and the technical methods they use to develop software for a product line. Two production strategies were reported in the work-ing group. One production strategy was based on a common platform, the other on a customizable federation of components.

The two strategies represent two end points on a scale of component customization. They differ according to the amount of variation incorporated in the architecture and components. They differ in the granularity of software that is visible to product management and in the separation of the asset group from the product development group.

In the platform production strategy[6] shown in Figure 3, a research and development (R&D) group develops a platform that is supplied at the start of each product development project. The platform consists of a tightly coupled configuration of standard components, and each project develops the variable portion of the product to specific customer requirements. The R&D group has the extensive technical expertise needed to develop common software. Each project, on the other hand, has extensive knowledge of the customer's unique requirements.

Because the platform is not very customizable, many versions of a product may result-each with small differences. Product features discovered over time to be common may be incorporated in the next release of the platform. Accordingly, this approach requires tight synchronization between the two groups; each project has to define what is included and excluded from the platform. The projects themselves share little software with each other. Because the platform is not very customizable, there is little entropy of software assets; however, it may not be sufficiently flexible for the market.

Companies would like to increase the leverage provided by a platform. Experience shows that a platform production strategy is at times cumbersome: either projects are unable to implement new features without significant effort, or the platform simply cannot accommodate the features desired by marketing [Cleaveland 96]. The challenge is to design for variability in addition to capitalizing on system commonalties. By incorporating variation in the architecture and components, the scope of products that can be built increases. Although up-front costs are greater, the downstream payoff may also be greater. Codifying variation extends the robustness of the architecture and components. The net effect is shown in Figure 3.

---

[6] Also referred to as a customer-teaming production strategy.

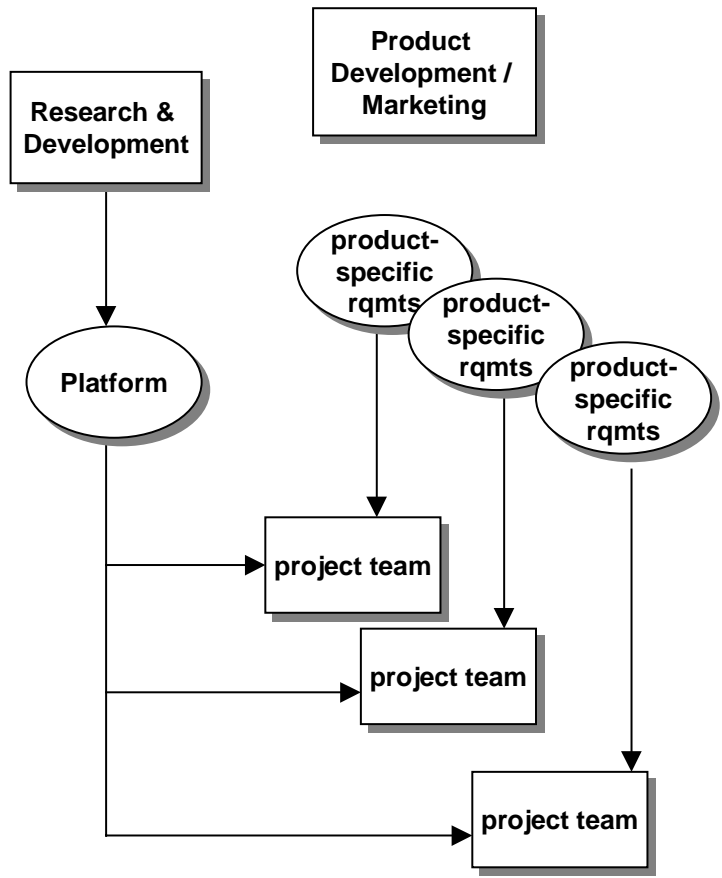

*Figure 3:    Platform Product Strategy*

In the customizable component product strategy shown in Figure 3, a software architecture team develops (or adapts) an architecture to accommodate the variety of products that are anticipated in the product line. Based on the architecture, product development is partitioned into component development (or acquisition) activities rather than projects. Members of the architecture team are also members of different component development projects. Unlike platform production strategy, the component development activities are visible to product line management. A baseline product is developed.

Customization to new product requirements is done at a subsystem (component) level. To the extent that the components are loosely coupled, multiple versions and configurations of components can be developed to build different products. Product integration is often a small task, possibly involving only linking executables.

This approach, however, requires a management infrastructure for managing complexity. Configuration management must track the requirements that are allocated to components in addition to the numerous configurations and variations that are developed. With concurrent development, there is a risk of architectural breakdown and loss of control, but there is also less dependency on internal product synchronization. The desired state is component-level customization, but with only a few allowable configurations.

For domains that are less defined and bounded, a customizable component production strategy seems to require the same group to develop the components and customize them for specific products. The component teams specialize in the technology; that is, the tools and solutions used to implement the components. Deep domain and product knowledge is needed to codify variation in the components, and the same knowledge is needed for customization. For well-defined and bounded domains, allowable component variation may be specified through parameters. In this case, members of the product integration team may specialize the component at integration, according to specific product requirements.

### 3.3.3 Conclusion

The working group was left with the following issues:

- When does it make sense to have two separate groups, one developing and sustaining an architecture and other assets, the other developing products using the architecture and assets? A two-group approach may be appropriate in green-field endeavors, where product development requires considerable research and development and deep product knowledge does not reside with product developers, but is this valid? Are there other situations?

- There is an inherent tension between the scope and the product-specific features covered by architecture and assets. Product managers want more product features to be provided, and senior managers want more products to be supported. How is this tension to be managed? Who decides on what goes into the architecture? How does one settle on the scope of architecture and the organizational level at which it is standardized?

# 4 Summary

The SEI's Second Product Line Practice Workshop explored the product line practices of technically sophisticated organizations with direct experience in software product lines. The presentations and discussions validated the pivotal pieces of the SEI's Product Line Practice Framework and suggested areas needing improvement and revision. The necessary practice areas identified at the previous product line workshop were underscored, and new areas were illuminated. Although terminology varied, the motivation to embrace a product line approach was consistently voiced. Moreover, key themes among successful product line endeavors emerged: long and deep domain experience, a legacy base from which to build, architectural excellence, and management commitment.

The working groups focused on the following specific practice areas within software engineering, technical management, and enterprise management: mining assets, domain analysis, metrics and data collection, configuration management, testing, planning, determining organizational structure, and determining a product line production strategy. In the latter two cases, the connections between the core assets, in particular the architecture, the production strategy, and organizational structure, were probed. The empirical and anecdotal evidence that the workshop participants brought to the discussion significantly enhanced our current understanding of the practices and issues. New issues were uncovered and many pervading ones remain unsolved. Still, a challenge for a product line approach is the repeatable integration of technical, business, and organizational practices.

We received feedback on the organization of the framework, the practices to include, and the need to show relationships among them. The need for continued exploration and codification of both technical and non-technical product line practices remains. In an effort to increase both the information base and the community interested in software product lines, the SEI intends to continue holding similar workshops and will also continue to report the workshop results to the software development community at large.

The information in this report will be incorporated into our framework, which will be refined and revised as the technology matures and as we continue to receive feedback and to work with the growing community of software engineers championing a product line approach. If you have any comments on this report and/or are using a product line approach in the development or acquisition of software-intensive systems and would like to participate in a future workshop, please send electronic mail to lmn@sei.cmu.edu.

# Glossary

**application engineering**
an engineering process that develops software products from partial solutions or knowledge embodied in software assets

**business model**
a framework that relates the different forms of a product line approach to an organization's business context and strategy

**domain**
an area of knowledge or activity characterized by a set of concepts and terminology understood by practitioners in that area

**domain analysis**
process for capturing and representing information about applications in a domain, specifically common characteristics and reasons for variability

**economies of scale**
the condition where fewer inputs such as effort and time are needed to produce greater quantities of a single output

**economies of scope**
the condition where fewer inputs such as effort and time are needed to produce a greater variety of outputs

Greater business value is achieved by jointly producing different outputs. Producing each output independently fails to leverage commonalities that affect costs. Economies of scope occur when it is less costly to combine two or more products in one production system than to produce them separately.

**investment analysis**
a process of estimating the value of an investment proposal to an organization

Investment analysis involves quantifying the costs and benefits of the investment, analyzing the uncertainties, and constructing a spending strategy. This analysis links the strategic and technical merits of an investment to its financial results.

**platform**
core software asset base that is reused across systems in the product line

**product family**
a group of systems built from a common set of assets

**product line**
a group of products sharing a common, managed set of features that satisfy needs of a selected market or mission area **product line approach**    a system of software production that uses software assets to modify, assemble, instantiate, or generate a line of software products

**product line architecture**    description of the structural properties for building a group of related systems (i.e., product line), typically the components and their interrelationships. The guidelines about the use of components must capture the means for handling variability discovered in the domain analysis or known to experts. Also called a reference architecture.

**product line system**    a member of a product line

**production system**    a system of people, functions, and assets organized to produce, distribute, and improve a family of products. Two functions included in the system are domain engineering and application engineering.

**software architecture**    structure or structures of the system, which is composed of software components, the externally visible properties of those components, and the relationships among them [Bass 98]

**system architectures**    software architecture plus execution and development environments

**software asset**    a description of a partial solution (such as a component or design document) or knowledge (such as a requirements database or test procedures) that engineers use to build or modify software products [Withey 96]

# REPORT DOCUMENTATION PAGE

| 1. AGENCY USE ONLY (LEAVE BLANK) | 2. REPORT DATE<br>April 1998 | 3. REPORT TYPE AND DATES COVERED<br>Final |
|---|---|---|

| 4. TITLE AND SUBTITLE<br><br>Second Product Line Practice Workshop Report | 5. FUNDING NUMBERS<br>C — F19628-95-C-0003 |
|---|---|

**6. AUTHOR(S)**

Len Bass, Gary Chastek, Paul Clements, Linda Northrop, Dennis Smith, James Withey

| 7. PERFORMING ORGANIZATION NAME(S) AND ADDRESS(ES)<br><br>Software Engineering Institute<br>Carnegie Mellon University<br>Pittsburgh, PA 15213 | 8. PERFORMING ORGANIZATION REPORT NUMBER<br><br>CMU/SEI-98-TR-015 |
|---|---|

| 9. SPONSORING/MONITORING AGENCY NAME(S) AND ADDRESS(ES)<br><br>HQ ESC/DIB<br>5 Eglin Street<br>Hanscom AFB, MA 01731-2116 | 10. SPONSORING/MONITORING AGENCY REPORT NUMBER<br><br>ESC-TR-98-015 |
|---|---|

**11. SUPPLEMENTARY NOTES**

| 12.A DISTRIBUTION/AVAILABILITY STATEMENT<br>Unclassified/Unlimited, DTIC, NTIS | 12.B DISTRIBUTION CODE |
|---|---|

**13. ABSTRACT (MAXIMUM 200 WORDS)**

The second Software Engineering Institute Product Line Practice Workshop was a hands-on meeting held in November 1997 to share industry practices in software product lines and to explore the technical and non-technical issues involved. This report synthesizes the workshop presentations and discussions, which identified factors involved in product line practices and analyzed issues in the areas of software engineering, technical management, and enterprise management.

| 14. SUBJECT TERMS<br>domain analysis, product family, product line practice, resue, software architecture, software platforms, software product lines | 15. NUMBER OF PAGES<br>32 |
|---|---|
| | 16. PRICE CODE |

| 17. SECURITY CLASSIFICATION OF REPORT<br>UNCLASSIFIED | 18. SECURITY CLASSIFICATION OF THIS PAGE<br>UNCLASSIFIED | 19. SECURITY CLASSIFICATION OF ABSTRACT<br>UNCLASSIFIED | 20. LIMITATION OF ABSTRACT<br>UL |
|---|---|---|---|